\documentclass[letterpaper, 10 pt, conference]{ieeeconf}  

\usepackage{xcolor,cite}
\usepackage{graphicx}
\usepackage{amsmath,physics,mathtools}
\usepackage{amssymb,bm,ulem}
\usepackage{amsthm}
\usepackage{algorithm,algorithmic}
\usepackage{comment}

\newcommand{\Initialize}[1][1]{
  \STATE \textbf{Initialize:}
}
\newtheorem{theorem}{Theorem}
\newtheorem{lemma}{Lemma}

\newtheorem {dfn}[theorem]{Definition}

\newcommand{\bo}[1]{\textbf{\textcolor{blue}{Bo: #1}}}

\IEEEoverridecommandlockouts            

\title{\LARGE \bf
Optimal Online Algorithms for One-Way Trading and Online Knapsack Problems: A Unified Competitive Analysis
}

\author{Ying Cao, Bo Sun and Danny H.K. Tsang
\thanks{This work was supported by the Hong Kong Research Grants Councils General Research Fund through Project 16211220. The authors are with Dept. of Electronic and Computer Engineering, the Hong Kong University of Science and Technology. Emails: \{ycaoan, bsunaa, eetsang\}@ust.hk}
}

\begin{document}

\maketitle
\thispagestyle{empty}
\pagestyle{empty}

\begin{abstract}
We study two canonical online optimization problems under capacity/budget constraints: the {\it fractional} one-way trading problem (OTP) and the {\it integral} online knapsack problem (OKP) under an infinitesimal assumption. Under the competitive analysis framework, it is well-known that both problems have the same optimal competitive ratio. 
However, these two problems are investigated by distinct approaches under separate contexts in the literature. 
There is a gap in understanding the connection between these two problems and the nature of their online algorithm design. 
This paper provides a unified framework for the online algorithm design, analysis and optimality proof for both problems. We find that the infinitesimal assumption of the OKP is the key that connects the OTP in the analysis of online algorithms and the construction of worst-case instances. 
With this unified understanding, our framework shows its potential for analyzing other extensions of OKP and OTP in a more systematic manner.  

\end{abstract}

\section{INTRODUCTION}
Online optimization under capacity/budget constraints is a classical and challenging problem. Two well-known examples are the one-way trading problem (OTP) and the online knapsack problem (OKP).

In the OTP, an investor plans to trade a total amount of 1 dollar into yen. The exchange rates $p_i$ arrive online and are bounded, i.e., $p_i\in [L,U]$, and the investor must immediately decide how much to trade at each exchange rate. If $x_i$ dollars are traded at the $i$th exchange rate $p_i$, $p_i x_i$ is the amount of yen the investor gains. Let $N$ denote the total number of exchange rates. The goal is to maximize the amount of yen traded after processing the $N$th exchange rate $\sum_{i=1}^N p_i x_i$, while respecting the budget limit $\sum_{i=1}^N x_i \le 1$. It is well-known that $(\ln (U/L)+1)$-competitive algorithms can be designed, e.g., the threat-based algorithm in \cite{el2001optimal} and the \textit{CR-Pursuit} algorithm in \cite{Lin2019}.

The 0-1 knapsack problem is a classic problem in computer science, where a decision-maker maximizes the total value of the items selected while the total weight does not exceed the normalized knapsack capacity limit of $1$. In the OKP, the items come one by one. The value $v_i$ and the weight $w_i$ of the $i$th item are only revealed upon its arrival. An online decision is made on whether to accept the item ($z_i=1$) or not ($z_i=0$). There exist no online algorithms with bounded competitive ratios for the OKP in the general setting \cite{babaioff2007knapsack}. However, ($\ln (U/L) +1$)-competitive algorithms can be designed \cite{zhou2008budget,Tan2020,zhang2017optimal} under the bounded value-to-weight ratio assumption (i.e., $v_i/w_i\in [L,U]$) and the {\it infinitesimal assumption} that the weight of each item is much smaller than the capacity (i.e., $\max_i w_i \ll 1$). The infinitesimal assumption is a technical simplification, but it has been shown to hold in practical applications such as cloud computing systems \cite{zhang2017optimal} and is widely accepted in the literature. In this paper, we make the same assumptions.

Both problems have appeared in numerous applications, including portfolio selection, cloud resource allocation, keyword auctions, etc. Thus, considerable attention has been paid to both problems and their many variants. Unbounded prices \cite{chin2015competitive} and interrelated prices \cite{schroeder2018optimal} have been considered for the OTP recently, and knapsacks with unknown capacity \cite{disser2017packing} and removable items \cite{han2013randomized} are interesting generalizations for the OKP. Additionally, different arrival models have been studied, such as stochastic arrivals \cite{tran2015efficient,fujiwara2011average} and random order \cite{albers2019improved}. In this paper, we make no assumptions on the arrival model.

Motivated by the gaps in the understanding of the nature of challenges in the online algorithm design, we aim to unify the online algorithms for the OTP and the OKP into one algorithmic framework, namely, a threshold-based algorithm, the competitive performance of which mainly depends on the threshold function. We provide a sufficient condition on the threshold function that can ensure a bounded competitive ratio, and design the best possible threshold function based on this sufficient condition. Finally, we derive the lower bound of the competitive ratios of the OTP and the OKP.
Although all results match those in the literature, the existing works approach the results by distinct methods and lack a systematic way of designing and analyzing related problems. This paper mainly focuses on the analysis and proofs rather than on the results. Our contributions are two-fold.
\begin{itemize}
    \item We unify the online algorithms for the OTP and the OKP into a threshold-based algorithm and show that the unified algorithm can achieve the optimal competitive ratios under a unified competitive analysis. 
    \item We provide new proofs for the lower bound of competitive ratios for the OTP and the OKP. The connection between these two problems is founded in the construction of the worst-case instances.
\end{itemize}

\section{A UNIFIED ALGORITHM and Our Results}
\subsection{Notations}
Since the two problems have originally distinct sets of terms, we unify the notations for the brevity of problem formulation and clarify the different meanings here. Let $x_i$ denote the amount of dollars traded at the $i$th exchange rate for the OTP, while for the OKP, it represents the capacity used after processing the $i$th item $w_i z_i$. Let $b_i$ denote the exchange rate (i.e., $p_i$) in the OTP and the value-to-weight ratio of the $i$th item in the OKP (i.e., ${v_i}/{w_i}$). The following optimization problem characterizes the offline problem of the OTP: 
\begin{equation}
\begin{aligned}
    \underset{\bm{x}}{\text{maximize}} \quad &\sum_{i=1}^{N}b_i x_i\\
    {\rm s.t.} \quad & \sum_{i=1}^{N}x_i \le 1, \\
    & x_i \ge 0, \forall i \in [N].
\end{aligned}
\tag{1}\label{primal_otp}
\end{equation}
The dual problem of (\ref{primal_otp}) is
\begin{equation}
\begin{aligned}
    \underset{\lambda}{\text{minimize}} \quad & \lambda\\
    {\rm s.t}. \quad & \lambda \geq b_i, \forall i \in [N].
\end{aligned}
\tag{2}\label{dual_otp}
\end{equation}
By changing the last constraint of (\ref{primal_otp}) to $0\le x_i\le w_i$, the resulting problem serves as an upper bound of the OKP due to the LP relaxation, and its dual is
\begin{equation}
\begin{aligned}
    \underset{\lambda,\bm{\beta}}{\text{minimize}} \quad &\lambda+\sum_{i=1}^{N}w_i \beta_i\\
    {\rm s.t.} \quad & \lambda+\beta_i\ge b_i,\forall i \in [N], \\
    & \lambda \ge 0,\beta_i \ge 0, \forall i \in [N],
\end{aligned}
\tag{3}\label{dual_okp}
\end{equation}
where $\lambda$ and $\beta_i$s are the dual variables of the corresponding dual programs.
\subsection{A Unified Algorithm} Both the OTP and the OKP target allocating one budget-constrained resource sequentially.
Since the current decision affects the future decisions through the budget constraint, we need an estimation of the value of the remaining resource to facilitate decision-making. We use a threshold function to estimate the value of resources.  

\begin{dfn}
A threshold function $\phi(y): [0,1]\to [0,\infty)$ estimates the marginal cost of a resource at utilization $y$. 
\end{dfn}

Denote the utilization level after the $i$th arrival by $y^{(i)}$. Given $\phi(y)$, we can estimate the pseudo-cost of allocating $x_i$ amount of resource by $\int_{y^{(i-1)}}^{y^{(i-1)} + x_i}\phi(\delta)d\delta$. Our unified algorithm then decides $x_i$ that maximizes the pseudo-revenue $b_ix_i - \int_{y^{(i-1)}}^{y^{(i-1)} + x_i}\phi(\delta)d\delta$.
The overall algorithm is summarized in Algorithm 1.
\begin{algorithm}
\caption{A Unified Algorithm}
\begin{algorithmic}
\Initialize $\phi(y)$, $y^{(0)} = 0$, $b_i\in[L,U]$;
\FOR{the $i$th time slot}
\STATE 1.
$
    x_i=\arg\max_{x\in \mathbb{S}} b_i x-\int_{y^{(i-1)}}^{y^{(i-1)}+x}\phi(\delta)d\delta.
$
\STATE 2. Update $y^{(i)} =  y^{(i-1)}+x_i$;
\STATE 3. If $y^{(i)}>1,x_i=0.$
\ENDFOR
\end{algorithmic}
\end{algorithm}
$\mathbb{S}$ denotes the set of all positive real numbers for the OTP and the set $\{0,w_i\}$ for the OKP, separately. Note that for the OKP, step 1 reduces to $x_i=\begin{cases}
w_i, & b_i\ge \phi(y^{(i-1)})\\
0, &  otherwise
\end{cases}$, which corresponds to the update equation in \cite{zhou2008budget}. Algorithm 1 can be easily applied in the posted-price setting by its nature.

\subsection{Main Results}
A standard measure for the performance of an online algorithm is the competitive ratio. Under the unified notation, define an arrival instance $\mathcal{A}$ as $\{b_i\}_{\forall i\in [N]}$ for the OTP, and as $\{b_i,w_i\}_{\forall i\in [N]}$ for the OKP. Denote the objective value achieved by the online algorithm and the offline optimal by $\text{ALG}(\mathcal{A})$ and $\text{OPT}(\mathcal{A})$, respectively, given the arrival instance $\mathcal{A}$. If
$
    \alpha = \max_{\mathcal{A}}{}\frac{\text{OPT}(\mathcal{A})}{\text{ALG}(\mathcal{A})}
$, then we say the online algorithm is $\alpha$-competitive. The competitive ratio of Algorithm 1 depends only on the choice of the function $\phi$. We find the sufficient conditions for $\phi$ for Algorithm 1 to be $\alpha$-competitive in the following theorem.

\begin{theorem}[Sufficiency]
\label{sufficiency}
Algorithm 1 is $\alpha$-competitive for both the OTP and the OKP if $\phi$ is given by
$$\phi(y) = \left\{
\begin{array}{cc}
    L & y \in [0,\omega]\\
    \varphi(y) & y \in [\omega,1]
\end{array},
\right .
$$
where $\omega$ is a budget/capacity utilization level that satisfies $\frac{1}{\alpha}\le \omega \le 1$, and $\varphi(y)$ is an increasing function that satisfies
\begin{equation}
\tag{4}\label{suffi_diffeq}
\left\{
\begin{array}{l}
    \varphi(y)\ge\frac{1}{\alpha}\varphi'(y),y \in [\omega,1] \\
    \varphi(\omega)=L, \varphi(1)\ge U.
\end{array}
\right . 
\end{equation}
\end{theorem}
In the theorem, $\phi$ is composed of two segments, one constant and the other exponential. Note that the functions used in \cite{zhou2008budget} and \cite{zhang2017optimal} satisfy the conditions. However, the basis of the functions is unknown; the authors do not explain the intuition behind the functions nor rigorously characterize the properties of the functions. In contrast, by the following theorem, we can characterize the performance limit over the space of all eligible functions and rigorously show the function that admits the smallest (best) competitive ratio.
\begin{theorem}
\label{opt_CR_for_alg1}
Given $L$ and $U$, the best competitive ratio that can be achieved by Algorithm 1 is $(\ln \theta+1)$, where $\theta=U/L$, and the corresponding $\phi^*$ is unique.
\end{theorem}
We show that no other online algorithms can perform better than Algorithm 1 using the following theorem.
\begin{theorem}
\label{lower_bound_thm}
Given $L$ and $U$, $(\ln\theta+1)$ is the lowest possible competitive ratio for both the OTP and the OKP.
\end{theorem}
In the next section, we introduce the primal-dual analysis framework, with which we prove Theorem \ref{sufficiency}. Subsequently, we prove Theorem \ref{opt_CR_for_alg1} by Gronwall's inequality. In Section \ref{lower_bounds}, we show Theorem \ref{lower_bound_thm} by adversarial arguments.

\section{COMPETITIVE ANALYSIS}
\subsection{Primal-Dual Competitive Analysis}
Given the arrival instance $\mathcal{A}$, we denote the primal and dual objective values after processing $b_n$ by $P_n(\mathcal{A})$ and $D_n(\mathcal{A})$, respectively. For simplicity, we drop the parenthesis and write $P_n$ and $D_n$ hereafter. We briefly introduce the framework by giving the following lemma.
\begin{lemma}
\label{primal-dual-lemma}
An online algorithm is $\alpha$-competitive if it can determine the primal variables $x$ and construct dual variables $\lambda$ based on the primal variables such that
\begin{itemize}
    \item (\textit{Feasible Solutions}) $x$ and $\lambda$ are feasible solutions of the primal and the dual,
    \item (\textit{Initial Inequality}) there exists an index $k\in [N]\cup \{0\}$ such that $P_k\ge \frac{1}{\alpha}D_k$,
    \item (\textit{Incremental Inequalities}) for $i\in \{k+1,\dots,N\}$,
    $$P_i-P_{i-1}\ge \frac{1}{\alpha}(D_i-D_{i-1}).$$
\end{itemize}
\end{lemma}

\begin{proof}
The primal feasibility is trivial since any online algorithm must first produce a feasible solution to the problem. It suffices to prove $P_N\ge \frac{1}{\alpha}D_N$ since
$$ \text{ALG} = P_N \ge \frac{1}{\alpha}D_N \overset{(a)}{\ge} \frac{1}{\alpha}D^* \overset{(b)}{\ge} \frac{1}{\alpha}\text{OPT},$$ where ($a$) is due to the dual feasibility, and ($b$) is due to the weak duality. Suppose there exists a $k$ such that $P_i-P_{i-1} \ge \frac{1}{\alpha}(D_i-D_{i-1})$ holds for all $i\in \{k+1,\dots,N\}$, then we have
$
    P_N-P_k \ge \frac{1}{\alpha}(D_N-D_k).
$
Combining this with the initial inequality leads to $P_N\ge \frac{1}{\alpha}D_N$. We thus complete the proof.
\end{proof}
Note that the primal-dual competitive analysis framework that we use is more general than those used in the existing works, in that the initial inequality starts from $k\in [N]\cup \{0\}$ rather than the original $0$. 

Next, we show the proofs of Theorems \ref{sufficiency} and \ref{opt_CR_for_alg1} for the OTP, and highlight the differences between them and those for the OKP.
\subsection{Analysis of OTP}

\begin{proof}[Proof of Theorem \ref{sufficiency}]
(\textbf{Feasible Solutions}) First we show that the primal and dual solutions given by Algorithm 1 are feasible: 
\begin{align}
    x_i = \begin{cases}
    \phi^{-1}(b_i) - y^{(i-1)} & b_i \ge \phi(y^{(i-1)})\\
    0 & b_i < \phi(y^{(i-1)})
\end{cases},
\tag{5}\label{primal_solution_expression}
\end{align}
where $\phi^{-1}(b)=\begin{cases}\omega & b=L \\ \varphi^{-1}(b) & b>L\end{cases}$. (\ref{primal_solution_expression}) ensures $\forall i,x_i \ge 0$, and $\phi(1)\ge U$ ensures $\phi^{-1}(U)\le 1$. Since $y^{(N)} = \phi^{-1}(\max_{i\in[N]} b_i)$, we have $y^{(N)}\le \phi^{-1}(U)\le 1$. Thus, the primal solutions are feasible. Construct the dual variables as
$
\lambda_i = \phi(y^{(i)}).
$ Since $\phi(y)$ is non-decreasing, $\lambda_N=\phi(y^{(N)}) \ge \phi(y^{(i)}), \forall i\in[N]$. Thus, $\lambda_N$ is a feasible solution to the dual.


(\textbf{Initial Inequality}) For the OTP, $P_0=0, D_0=\phi(y^{(0)})=L>0$. When $k\ge 1$, the primal objective at the end of the $k$th time slot is
$ P_k=\sum_{i=1}^k b_i x_i, $
while the dual objective is
$ D_k=\lambda_k=\phi(y^{(k)}). $

Since $y^{(0)}=0$ and $\phi(0) = L\le b_i,\forall i$, by (\ref{primal_solution_expression}), we have
$$x_1 = \phi^{-1}(b_1) = \begin{cases}\omega & b_1=L \\ \varphi^{-1}(b_1) & b_1>L\end{cases}.$$ Because $\varphi(y)$ is an increasing function, we have $x_1\ge \omega\ge \frac{1}{\alpha}.$ Since $b_1=\phi(x_1)$, it follows that $$ P_1= b_1 x_1 \ge  \frac{b_1}{\alpha}=\frac{1}{\alpha}\phi(x_1)=\frac{1}{\alpha}D_1.$$

(\textbf{Incremental Inequalities})
Next we show the incremental inequalities for $i>1$. Note that when $x_i=0$, $P_i=P_{i-1}$ and $D_i=D_{i-1}$. In this case, the incremental inequality $P_i-P_{i-1}\ge \frac{1}{\alpha}(D_i-D_{i-1})$ always holds. Thus, we only need to focus on the case where $x_i>0,\forall i>1$, when the behavior of the algorithm is controlled by the second segment of $\phi$, which satisfies $\varphi(y)\ge\frac{1}{\alpha}\varphi'(y)$ for $y\in [\omega,1]$ and two boundary conditions $\varphi(\omega)=L$ and $\varphi(1)\ge U$.
 
 The change in the primal objective is given as follows:
 $$ P_i-P_{i-1}=b_i x_i\overset{(a)}{=} \phi(y^{(i)})x_i,$$
where ($a$) is due to (\ref{primal_solution_expression}) and $y^{(i)}=y^{(i-1)}+x_i$.

The change in the dual objective is given as follows:
$
    D_i-D_{i-1}= \varphi(y^{(i)})-\varphi(y^{(i-1)}).
$ By the Cauchy mean value theorem, for every segment $[y^{(i-1)},y^{(i)}]$, there exists a $\delta_i \in [y^{(i-1)},y^{(i)}]$ such that

$$ \frac{\varphi(y^{(i)})-\varphi(y^{(i-1)})}{y^{(i)}-y^{(i-1)}}=\varphi'(\delta_i).$$
Since $\forall y\in [\omega,1],\varphi(y)\ge\frac{1}{\alpha}\varphi'(y)$, and $\varphi(y)$ is increasing, we have $\alpha \varphi(y^{(i)}) \ge \alpha \varphi(\delta_i) \ge \frac{\varphi(y^{(i)})-\varphi(y^{(i-1)})}{y^{(i)}-y^{(i-1)}}. $
Because $y^{(i)}-y^{(i-1)}>0$, we have
$ \varphi(y^{(i)})(y^{(i)}-y^{(i-1)})\ge \frac{1}{\alpha}(\varphi(y^{(i)})-\varphi(y^{(i-1)})),$ where the LHS is $P_i-P_{i-1}$, and the RHS is $\frac{1}{\alpha}(D_i-D_{i-1})$. Thus, $P_i-P_{i-1}\ge \frac{1}{\alpha}(D_i-D_{i-1})$ holds for all $i>1$. 

Therefore, Theorem \ref{sufficiency} holds for the OTP.
\end{proof}

Theorem \ref{opt_CR_for_alg1} characterizes the performance limit of Algorithm 1.

\begin{proof}[Proof of Theorem \ref{opt_CR_for_alg1}]
(\textbf{Best Competitive Ratio}) By the differential form of Gronwall's Inequality \cite{mitrinovic1991inequalities}, if there exists a $\varphi$ that satisfies $\varphi(y)\ge\frac{1}{\alpha}\varphi'(y),y \in [\omega,1],$ where $\omega\in[\frac{1}{\alpha},1]$, it is bounded as follows:
$$\varphi(y)\le \varphi(\omega)\exp{\bigg(\int_{\omega}^{y}\alpha dt\bigg)}, y\in[\omega,1].$$
Substituting the first boundary condition $\varphi(\omega)=L$, we have
$\varphi(y)\le L\exp{\big(\alpha(y-\omega)\big)}, y\in[\omega,1].$ If the other boundary condition $\varphi(1)\ge U$ holds, it implies
$
L\exp{\big(\alpha(1-\omega)\big)}\ge U,
$ otherwise $\varphi(1)\le L\exp{\big(\alpha(1-\omega)\big)} < U$, which incurs infeasibility. From the inequality above, we have
$
\omega \le 1-\frac{1}{\alpha}\ln \theta.
$
A necessary condition for $\omega\ge \frac{1}{\alpha}$ to hold is
$
1-\frac{1}{\alpha}\ln \theta \ge \frac{1}{\alpha},
$
and thus the competitive ratio
$\alpha \ge \ln{\theta}+1.$

(\textbf{$\Phi^*$ and Its Uniqueness}) When $\alpha$ takes the smallest possible $\alpha^*=\ln\theta+1$, the corresponding  $\phi^*$s satisfy
$$\phi^*(y) = \left\{
\begin{array}{cc}
    L & y \in [0,\omega]\\
    \varphi^*(y) & y \in [\omega,1]
\end{array},
\right .
$$ where $\omega \in [\frac{1}{\ln \theta+1},1]$ and $\varphi^*$s are given by
\begin{equation}
\tag{6}\label{opt_diffeq}
\left\{
\begin{array}{l}
    \varphi^*(y)\ge\frac{1}{\ln \theta+1}{\varphi^*}'(y),y \in [\omega,1] \\
    \varphi^*(\omega)=L, \varphi^*(1)\ge U.
\end{array}
\right . 
\end{equation}
By Gronwall's inequality, we have
\begin{align*}
    \tag{7} \label{gronwall_ineq_opt}
    \varphi^*(y)&\le L\exp\big((\ln\theta+1)(y-\omega)\big) \\
    &\overset{(a)}{\le} L \exp\big((\ln\theta+1)y-1\big),y\in[\omega,1],
\end{align*}
where ($a$) is due to $\omega\ge \frac{1}{\ln\theta+1}$. Then we have
$
    \varphi^*(1)\le L\exp(\ln\theta)=L\theta=U.
$
Combining with the second boundary condition $\varphi^*(1)\ge U$, we have $\varphi^*(1)= U$. Substituting this into (\ref{gronwall_ineq_opt}), we have
$
    \omega \le \frac{1}{\ln\theta+1}.
$
Because $\omega\ge \frac{1}{\ln\theta+1}$ as stated in Theorem \ref{sufficiency}, we have $\omega=\frac{1}{\ln\theta+1}$. Therefore, the solution space of (\ref{opt_diffeq}) is equivalent to the solution space of the following differential inequality with equality boundary conditions:
\begin{equation}
\tag{8}\label{opt_final_diffineq}
\left\{
\begin{array}{l}
    u(y)\ge\frac{1}{\ln \theta+1}{u}'(y),y \in [\frac{1}{\ln\theta+1},1] \\
    u(\frac{1}{\ln\theta+1})=L, u(1)=U.
\end{array}
\right . 
\end{equation}
The differential equation counterpart is as follows:
\begin{equation}
\tag{9}\label{opt_final_diffeq}
\left\{
\begin{array}{l}
    v(y)=\frac{1}{\ln \theta+1}{v}'(y),y \in [\frac{1}{\ln\theta+1},1] \\
    v(\frac{1}{\ln\theta+1})=L, v(1)=U.
\end{array}
\right .
\end{equation}
The unique solution to (\ref{opt_final_diffeq}) is $v(y)=\frac{L}{e}e^{(\ln\theta+1)y}$. Suppose $u$ is a feasible solution to (\ref{opt_final_diffineq}), by Gronwall's inequality, $u(y)\le v(y), \forall y\in [\frac{1}{\ln\theta+1},1]$. Next, we are going to show that the solution of (\ref{opt_final_diffineq}) is unique and is exactly $v(y)$.

Suppose $u(y)<v(y)$ for $y\in \mathbb{I}$, where $\mathbb{I}\subset [\frac{1}{\ln\theta+1},1]$. We know that for any $y\in [\frac{1}{\ln\theta+1},1]$,
$
    v'(y)=(\ln\theta+1)v(y).
$
Thus, for $y\in \mathbb{I}$, we have
$$v'(y)=(\ln\theta+1)v(y)>(\ln\theta+1)u(y)\ge u'(y).$$
Take the integral of $u'$ over $[\frac{1}{\ln\theta+1},1]$, we have
$
    \int_{\frac{1}{\ln\theta+1}}^{1}u'(y)dy=u\big|_{\frac{1}{\ln\theta+1}}^{1}=U-L,
$
which can also be expressed as
\begin{align*}
    \int_{\frac{1}{\ln\theta+1}}^{1}u'(y)dy&=\int_{\mathbb{I}}u'(y)dy+\int_{[\frac{1}{\ln\theta+1},1]\backslash\mathbb{I}} u'(y)dy\\
    &<\int_{\mathbb{I}}v'(y)dy+\int_{[\frac{1}{\ln\theta+1},1]\backslash\mathbb{I}} u'(y)dy\\
    &=\int_{\frac{1}{\ln\theta+1}}^{1}v'(y)dy=v\big|_{\frac{1}{\ln\theta+1}}^{1}=U-L,
\end{align*}
which shows $U-L<U-L$. Thus, $u(y)=v(y)$ for $y\in [\frac{1}{\ln\theta+1},1]$. In conclusion, the optimal $\phi^*$ achieving competitive ratio $(\ln\theta+1)$ is unique, and $$\phi^*(y) = \left\{
\begin{array}{cc}
    L & y \in [0,\frac{1}{\ln\theta+1}]\\
    \frac{L}{e}e^{(\ln\theta+1)y} & y \in (\frac{1}{\ln\theta+1},1]
\end{array}.
\right .
$$
\end{proof}
\subsection{Analysis of OKP}
We highlight the differences in the analysis of the OKP. The primal feasibility holds trivially and the dual variables are constructed as follows:
\begin{align*}
    \lambda = \lambda_N,\quad \beta_i=\begin{cases}b_i-\lambda_i & x_i = w_i\\
    0 & x_i  = 0\end{cases},
\end{align*}
where $\lambda_i = \phi(y^{(i-1)})$. When $x_i = w_i$, based on the decision-making rule (Step 1 in the algorithm), we must have $b_i \ge \phi(y^{(i-1)})$. Therefore, $\beta_i \ge 0, \forall i \in [N]$. The constraint of the dual problem is
\begin{align*}
 \lambda + \beta_i - b_i = 
 \begin{cases}
 \lambda - \lambda_i \ge 0& x_i = w_i\\
 \lambda - b_i \ge 0 & x_i = 0
 \end{cases}.
\end{align*}
Thus, the dual feasibility holds.

Assume that the online algorithm will accept the first $k$ items, and $\omega=\sum_{i=1}^k w_i $. Also note that $\lambda_i = L, \forall i\in [k]$. Then we have
\begin{align*}
 D_k &= \lambda_k + \sum_{i=1}^k w_i \beta_i = \lambda_k + \sum_{i=1}^k w_i (b_i - \lambda_i)\\
 &= L(1 - \omega) + \sum_{i=1}^k w_ib_i\\
 &\le \frac{1}{\omega}(\sum_{i=1}^k w_ib_i) \le \alpha \sum_{i=1}^k w_ib_i = \alpha P_k.
\end{align*}
Thus, there exists $k$ that satisfies the initial inequality.

With regard to the incremental inequalities, we have
$
    P_i-P_{i-1} = b_i w_i,
    D_i-D_{i-1} = \varphi(y^{(i)})-\varphi(y^{(i-1)})+w_i(b_i-\varphi(y^{(i-1)})
    \overset{(a)}{=}w_i [\varphi'(y^{(i-1)})+b_i-\varphi(y^{(i-1)})],
$
where $(a)$ is due to the fact $\varphi'(y^{(i-1)})=\frac{\varphi(y^{(i)})-\varphi(y^{(i-1)})}{w_i}$ and $w_i=y^{(i)}-y^{(i-1)}$ (using the infinitesimal weight assumption). Combining the ODE (\ref{suffi_diffeq}) with the fact that $b_i\ge \phi(y^{(i-1)}) = \varphi(y^{(i-1)})$, the incremental inequality holds for $i\in [N]$. Thus, Theorem $\ref{sufficiency}$ holds for the OKP. Note that the proof of Theorem \ref{opt_CR_for_alg1} holds generally for the two problems.
\section{Lower Bounds}
\label{lower_bounds}
In this section, we show that the lower bound of the OTP and that of the OKP coincide. Denote Algorithm 1 with $\phi^*$ by $\text{ALG}_{\phi^*}$. We first present the proofs for the OTP, then call attention to the differences for the OKP case.

\subsection{Lower Bounds of OTP}
 Below we find the family of the worst-case sequences under which $\text{ALG}_{\phi^*}$ incurs a ratio of $\ln\theta+1$.

\begin{lemma}
\label{worst-case-seq}
Given $L$ and $U$, the family of the worst-case sequences of $\text{ALG}_{\phi^*}$ in the OTP are denoted by $\{\hat{\delta}_k\}_{k\in \mathbb{N}^+}$, where $\hat{\delta}_k=\{\hat{b}_1,\dots,\hat{b}_k\}$, $\hat{b}_i\in[L,U]$ and the rates satisfy
$$
\hat{b}_1=L,\hat{b}_{i}=\hat{b}_{i-1}+\epsilon_{i-1}, i>1, \lim_{k\rightarrow \infty} \hat{b}_k=U,
$$
where $\epsilon_i$s are infinitesimal positive values. The amount traded by $\text{ALG}_{\phi^*}$ at the exchange rate $\hat{b}_i$ is denoted by $\hat{x}_i$, which satisfies
$$
\hat{x}_1=\frac{1}{\ln\theta+1}, \hat{x}_i=\frac{\ln \hat{b}_i/\hat{b}_{i-1}}{\ln\theta+1},i>1,
\lim_{k\rightarrow \infty} \sum_{i=1}^k \hat{x}_i = 1.
$$
\end{lemma}
\begin{proof}
The proof of Lemma \ref{worst-case-seq} is in the Appendix.
\end{proof}

\begin{proof}[Proof of Theorem \ref{lower_bound_thm}]
Let ALG be any online algorithm different from $\text{ALG}_{\phi^*}.$ We show that ALG cannot achieve a competitive ratio smaller than $\ln\theta+1$ by using an adversarial argument.

Let $\hat{\delta}=\{L,L+\epsilon_1,\dots,U\}$. First present $\hat{b}_1=L$ to ALG. If ALG exchanges $x'_1<\hat{x}_1=1/(\ln\theta+1)$, then we end the sequence, and on doing so, ALG cannot achieve $\ln\theta+1$, because $\frac{\text{OPT}(\hat{\delta}_1)}{\text{ALG}(\hat{\delta}_1)}=\frac{1}{x'_1}>\ln\theta+1.$ Thus, we can assume that ALG spends an amount $x'_1\ge\hat{x}_1$, in this case, we continue and present $\hat{b}_2$ to ALG. In general, if after processing the $k$th exchange rate, the total amount of dollars spent is less than $\sum_{i=1}^k \hat{x}_i$, we immediately end the sequence. Otherwise, we continue to present $\hat{b}_{k+1}$, etc. 

Let $f(k)=\sum_{i=1}^k (x'_i-\hat{x}_i)$. Let $\mathbb{K}=\{k\in \mathbb{N}|f(k)<0\}$, denote the minimum in $\mathbb{K}$ as $j$, we have
\begin{align*}
    &x'_1 \ge \hat{x}_1, \\
    &x'_1+x'_2 \ge \hat{x}_1+\hat{x}_2, \dotsc \\
    &\sum_{i=1}^{j-1}x'_i \ge \sum_{i=1}^{j-1}\hat{x}_i, \\
    &\sum_{i=1}^{j}x'_i < \sum_{i=1}^{j}\hat{x}_i.
\end{align*}
Thus, ALG can gain more by spending exactly the same as $\text{ALG}_{\phi^*}$ at the first $(j-1)$ exchange rates and by spending $\Tilde{x'_j}=x'_j+\sum_{i=1}^{j-1}(x'_i-\hat{x}_i)$ at the $j$th exchange rate. Since $\Tilde{x'_j}<\hat{x}_j$, ALG cannot guarantee the competitive ratio of $\ln\theta+1$. If $f(k)\ge0$ for all $k\in \mathbb{N}^+$, we have
\begin{align*}
    \liminf_{k\rightarrow\infty} f(k)\ge0,\limsup_{k\rightarrow\infty} f(k)\ge0.
\end{align*}
Since ALG cannot exceed the capacity limit, we have $\lim_{k\rightarrow\infty}\sum_{i=1}^k x'_i\le 1,$
and we also have $\lim_{k\rightarrow \infty} \sum_{i=1}^k \hat{x}_k = 1$, therefore, we have 
$\lim_{k\rightarrow\infty}f(k)\le0.$
For an infinite sequence $f(k)$, the limit exists iff 
$$\limsup f(k)=\liminf f(k)=\lim f(k),$$ so we have $ \lim_{k\rightarrow\infty}f(k)= 0. $ By the Abel transformation, we have
$
    \sum_{i=1}^k \hat{b}_i (x'_i-\hat{x}_i)= \sum_{i=1}^{k-1}f(i)(\hat{b}_i-\hat{b}_{i+1})+f(k)\hat{b}_k \\
    \overset{(a)}{\le} f(k)\hat{b}_k,
$
where ($a$) is due to the monotonicity of $\{\hat{b}_i\}$.

Thus, the performance gap between ALG and $\text{ALG}_{\phi^*}$ for this infinite exchange rate sequence is
$
    \lim_{k\rightarrow\infty}\sum_{i=1}^k \hat{b}_i (x'_i-\hat{x}_i) \le \lim_{k\rightarrow\infty}f(k)\hat{b}_k=0.
$

Therefore, any online algorithm for the OTP cannot achieve a better competitive ratio than $\text{ALG}_{\phi^*}$. The lowest possible competitive ratio is $\ln\theta+1$.

\end{proof}

\subsection{Lower Bounds of OKP}
We show that with a slight modification, $\{\hat{\delta}_k\}_{k\in \mathbb{N}^+}$ are also the worst-case sequences for the OKP.

Consider a family of the value-to-weight ratio sequences $\{I_b\}$ indexed by $b\in[L,U]$. $I_b$ is composed of a continuum of subsequences, with the ratios in the $i$th subsequence all being $\hat{b}_i,i\in \mathbb{N}^+$, where $\hat{b}_i\le b$, which is given in Lemma \ref{worst-case-seq}. The length of each subsequence is sufficiently large so that it can fulfill the capacity of the knapsack even if it is presented alone. Note that given $I_b$, the resource allocation strategy is analogous to the OTP case. The offline optimal solution is to only select from the subsequence with $\hat{b}_i=b$ until reaching the capacity limit, whereas $\text{ALG}_{\phi^*}$ will select a value-to-weight ratio as long as it is no less than $\phi^*(y)$, where $y$ is the current capacity utilization level. Therefore, $\{I_b\}_{b\in [L,U]}$ are the worst-case sequences for the OKP.

With regards to the proof of Theorem \ref{lower_bound_thm}, one can replace the worst sequence $\hat{\delta}$ with $I_U$, present a subsequence instead of an arrival at a time to $\text{ALG}$, and act adversarially in the same way in response to the decisions made by $\text{ALG}$, and the results still hold.

\section{CONCLUSION}
We provide a unified threshold-based algorithm for two well-known online problems, namely, the one-way trading problem and the online knapsack problem. We show the sufficient conditions for the algorithm to have a bounded competitive ratio for both problems via a unified competitive analysis. We reveal the threshold function that can achieve the best (smallest) competitive ratio, show that it matches the lower bounds, and present new proofs for the lower bounds for both problems. This is the first work that introduces the connections between the OTP and the OKP and provides a unified algorithmic framework for both of them, and we believe there is much more to be explored in this direction, i.e., unifying the online algorithm design.


\section*{APPENDIX}
\begin{proof}[Proof of Lemma \ref{worst-case-seq}]
Denote any strictly-increasing sequence with length $k$ by $\delta_k=\{b_1,\dots,b_k\}$. We can simply focus on the strictly-increasing sequences, because $\text{ALG}_{\phi^*}$ only trades something when the current exchange rate is the new high observed. Any normal sequences can be transformed into a strictly-increasing sequence by keeping the exchange rate higher than all of its predecessors and omitting the rest, and the optimal in hindsight is not affected by this transformation. By (\ref{primal_solution_expression}), we have
\begin{align*}
    x_1&={\phi^*}^{-1}(b_1)
    =\frac{\ln(b_1 e/L)}{\ln\theta+1},\\
    x_i&={\phi^*}^{-1}(b_i)-{\phi^*}^{-1}(b_{i-1})\\
    &=\frac{\ln(b_i/b_{i-1})}{\ln\theta+1},i\ge 2.
\end{align*}
Denote the total amount of yen $\text{ALG}_{\phi^*}$ trades for $\delta_k$ by ALG($\delta_k$) and the offline optimal amount by OPT($\delta_k$). We have $\text{OPT}(\delta_k)=b_k,$ and
\begin{align*}
    \text{ALG}(\delta_k)=\sum_{i=1}^k b_i x_i=\frac{b_1 \ln(b_1 e/L)+\sum_{i=2}^k b_i \ln(b_i/b_{i-1})}{\ln\theta+1}.
\end{align*}
Let $r_k(b_1,\dots,b_k)=\frac{b_1 \ln(b_1 e/L)+\sum_{i=2}^k b_i \ln(b_i/b_{i-1})}{b_k}$. Thus, the competitive ratio for $\text{ALG}_{\phi^*}$ can be expressed as
\begin{align*}
    \underset{\{b_1,\dots,b_k,k\}}{\max}\frac{\text{OPT}(\delta_k)}{\text{ALG}(\delta_k)}
    &= \frac{\ln\theta+1}{\underset{\{b_1,\dots,b_k,k\}}{\min}r_k(b_1,\dots,b_k)}.
\end{align*}
Because $\text{ALG}_{\phi^*}$ can achieve $\ln\theta+1$ with $\phi^*$ by Theorem \ref{opt_CR_for_alg1}, we know that $\underset{\{b_1,\dots,b_k,k\}}{\min}r_k=1$. Next, we look for $\{b_1,\dots,b_k\} $ that minimize $r_k(b_1,\dots,b_k)$ for each $k$. When $k=1$, $r_1(b_1)=\ln(b_1e/L)\ge 1$, therefore, $\hat{\delta}_1=\{L\}$, $\hat{x}_1=\frac{1}{\ln\theta+1}$ and $\frac{\text{OPT}(\hat{\delta}_1)}{\text{ALG}(\hat{\delta}_1)}=\ln\theta+1$. When $k=2$,
\begin{align*} 
    r_2(b_1,b_2)=\frac{b_1 \ln(b_1 e/L)+b_2 \ln(b_2/b_{1})}{b_2}.
\end{align*}
The first order derivatives are
\begin{align*}
     \frac{\partial r_2}{\partial b_1}&=\frac{\ln (b_1 e/L)+1-b_2/b_1}{b_2},\\
     \frac{\partial r_2}{\partial b_2}&=\frac{b_2-b_1\ln(b_1 e/L)}{{b_2}^2}.
\end{align*}

We notice that $\partial r_2/\partial b_1$ and $\partial r_2/\partial b_2$ cannot be zero simultaneously. This means that $r_2$ has no critical points, and the minimum value of $r_2$ on $[L,U]\times[L,U]$ must be on one of the four boundary points. It turns out that $r_2$ reaches minimum when $(b_1,b_2)=(L,L)$. We need to find a close neighbor to $(L,L)$ with $b_2>b_1$ and whose value does not increase too much. Notice that $\partial r_2/\partial b_1\big|_{(L,L)}>0,\partial r_2/\partial b_2\big|_{(L,L)}=0$, increasing $b_2$ to $b_2+\epsilon$ with infinitesimal positive $\epsilon$ should incur the least inaccuracy, thus, $\hat{\delta}_2=\{L,L+\epsilon\}$ and $\frac{\text{OPT}(\hat{\delta}_2)}{\text{ALG}(\hat{\delta}_2)}\rightarrow(\ln\theta+1)^-$ as $\epsilon\rightarrow 0^+$.
For general $k\ge 3$,
\begin{align*}
    r_k(b_1,\dots,b_k)=\frac{b_1\ln(b_1 e/L)+\sum_{i=2}^k b_i\ln(b_i/b_{i-1})}{b_k}.
\end{align*}
The first order derivatives are:
\begin{align*}
    \frac{\partial r_k}{\partial b_1}=\frac{\ln (b_1 e/L)+1-b_2/b_1}{b_2},
\end{align*}
\begin{align*}
    \frac{\partial r_k}{\partial b_k}=\frac{b_k-b_{k-1}r_{k-1}(b_1,\dots,b_{k-1})}{{b_k}^{2}},
\end{align*}
\begin{align*}
    \frac{\partial r_k}{\partial b_i}=\frac{\ln(b_i/b_{i-1})+1-b_{i+1}/b_i}{b_k}, i=2,\dots,k-1.
\end{align*}
A commonality is that, $\partial r_k/\partial b_k$ and $\partial r_k/\partial b_{k-1}$ cannot be zero at the same time, and $r_k$ reaches minimum when $b_i=L,i\in [k]$. The increasing sequence closest to the minimum point is $
\{L,L+\epsilon_1,\dots,L+\sum_{i=1}^{k-1}\epsilon_i\}$, where $\epsilon_i$s are infinitesimal positive values, and we have $\frac{\text{OPT}(\hat{\delta}_k)}{\text{ALG}(\hat{\delta}_k)}\rightarrow(\ln\theta+1)^-$ as $\sum_{i=1}^{k-1} \epsilon_i\rightarrow 0^+.$ Actually, each $\hat{\delta}_k$ is the prefix of $\hat{\delta}_m,m\ge k.$ From these observations, we claim that as long as the exchange rate sequence increases slowly enough from $L$, it is the worst-case sequence for $\text{ALG}_{\phi^*}$.

To verify the claim, let $\hat{b}_k$ be $L+\sum_{i=1}^{k-1}\epsilon_i$, we have
\begin{align*}
    \text{ALG}(\hat{\delta}_k)&=\sum_{i=1}^k \hat{b}_i x_i \\
    &= \frac{L}{\ln\theta+1} + \sum_{i=2}^k \hat{b}_i \bigg(\frac{\ln(\hat{b}_i)-\ln(\hat{b}_{i-1})}{\ln\theta+1}\bigg) \\
    &= \frac{L}{\ln\theta+1}+\int_{L}^{\hat{b}_k}\gamma\cdot\frac{d\ln(\gamma)}{\ln\theta+1} 
    = \frac{\hat{b}_k}{\ln\theta+1},
\end{align*}
and thus 
$$\frac{\text{OPT}(\hat{\delta}_k)}{\text{ALG}(\hat{\delta}_k)}=\underset{\{b_1,\dots,b_k,k\}}{\max}\frac{\text{OPT}(\delta_k)}{\text{ALG}(\delta_k)}=\ln\theta+1.$$
Since the exchange rate is upper bounded by $U$, by the monotone convergence theorem, we have
$$ \lim_{k\rightarrow\infty}\hat{b}_k = U, $$ and thus $\lim_{k\rightarrow \infty} \sum_{i=1}^k \hat{x}_i = 1$.
\end{proof}


\bibliographystyle{IEEEtran}
\normalem
\bibliography{my_bib}

\begin{thebibliography}{10}
\providecommand{\url}[1]{#1}
\csname url@rmstyle\endcsname
\providecommand{\newblock}{\relax}
\providecommand{\bibinfo}[2]{#2}
\providecommand\BIBentrySTDinterwordspacing{\spaceskip=0pt\relax}
\providecommand\BIBentryALTinterwordstretchfactor{4}
\providecommand\BIBentryALTinterwordspacing{\spaceskip=\fontdimen2\font plus
\BIBentryALTinterwordstretchfactor\fontdimen3\font minus
  \fontdimen4\font\relax}
\providecommand\BIBforeignlanguage[2]{{%
\expandafter\ifx\csname l@#1\endcsname\relax
\typeout{** WARNING: IEEEtran.bst: No hyphenation pattern has been}%
\typeout{** loaded for the language `#1'. Using the pattern for}%
\typeout{** the default language instead.}%
\else
\language=\csname l@#1\endcsname
\fi
#2}}

\bibitem{el2001optimal}
R.~El-Yaniv, A.~Fiat, R.~M. Karp, and G.~Turpin, ``Optimal search and one-way
  trading online algorithms,'' \emph{Algorithmica}, vol.~30, no.~1, pp.
  101--139, 2001.

\bibitem{Lin2019}
Q.~Lin, H.~Yi, J.~Pang, M.~Chen, A.~Wierman, M.~Honig, and Y.~Xiao,
  ``Competitive online optimization under inventory constraints,''
  \emph{Proceedings of the ACM on Measurement and Analysis of Computing
  Systems}, vol.~3, no.~1, pp. 1--28, 2019.

\bibitem{babaioff2007knapsack}
M.~Babaioff, N.~Immorlica, D.~Kempe, and R.~Kleinberg, ``A knapsack secretary
  problem with applications,'' in \emph{Approximation, randomization, and
  combinatorial optimization. Algorithms and techniques}.\hskip 1em plus 0.5em
  minus 0.4em\relax Springer, 2007, pp. 16--28.

\bibitem{zhou2008budget}
Y.~Zhou, D.~Chakrabarty, and R.~Lukose, ``Budget constrained bidding in keyword
  auctions and online knapsack problems,'' in \emph{International Workshop on
  Internet and Network Economics}.\hskip 1em plus 0.5em minus 0.4em\relax
  Springer, 2008, pp. 566--576.

\bibitem{Tan2020}
\BIBentryALTinterwordspacing
X.~Tan, B.~Sun, A.~Leon-Garcia, Y.~Wu, and D.~H. Tsang, ``Mechanism design for
  online resource allocation: A unified approach,'' \emph{Proc. ACM Meas. Anal.
  Comput. Syst.}, vol.~4, no.~2, June 2020. [Online]. Available:
  \url{https://doi.org/10.1145/3392142}
\BIBentrySTDinterwordspacing

\bibitem{zhang2017optimal}
Z.~Zhang, Z.~Li, and C.~Wu, ``Optimal posted prices for online cloud resource
  allocation,'' \emph{Proceedings of the ACM on Measurement and Analysis of
  Computing Systems}, vol.~1, no.~1, pp. 1--26, 2017.

\bibitem{chin2015competitive}
F.~Y. Chin, B.~Fu, J.~Guo, S.~Han, J.~Hu, M.~Jiang, G.~Lin, H.-F. Ting,
  L.~Zhang, Y.~Zhang, \emph{et~al.}, ``Competitive algorithms for unbounded
  one-way trading,'' \emph{Theoretical Computer Science}, vol. 607, pp. 35--48,
  2015.

\bibitem{schroeder2018optimal}
P.~Schroeder, R.~Dochow, and G.~Schmidt, ``Optimal solutions for the online
  time series search and one-way trading problem with interrelated prices and a
  profit function,'' \emph{Computers \& Industrial Engineering}, vol. 119, pp.
  465--471, 2018.

\bibitem{disser2017packing}
Y.~Disser, M.~Klimm, N.~Megow, and S.~Stiller, ``Packing a knapsack of unknown
  capacity,'' \emph{SIAM Journal on Discrete Mathematics}, vol.~31, no.~3, pp.
  1477--1497, 2017.

\bibitem{han2013randomized}
X.~Han, Y.~Kawase, and K.~Makino, ``Randomized algorithms for removable online
  knapsack problems,'' in \emph{Frontiers in Algorithmics and Algorithmic
  Aspects in Information and Management}.\hskip 1em plus 0.5em minus
  0.4em\relax Springer, 2013, pp. 60--71.

\bibitem{tran2015efficient}
L.~Tran-Thanh, Y.~Xia, T.~Qin, and N.~R. Jennings, ``Efficient algorithms with
  performance guarantees for the stochastic multiple-choice knapsack problem,''
  in \emph{Twenty-Fourth International Joint Conference on Artificial
  Intelligence}, 2015.

\bibitem{fujiwara2011average}
H.~Fujiwara, K.~Iwama, and Y.~Sekiguchi, ``Average-case competitive analyses
  for one-way trading,'' \emph{Journal of Combinatorial Optimization}, vol.~21,
  no.~1, pp. 83--107, 2011.

\bibitem{albers2019improved}
S.~Albers, A.~Khan, and L.~Ladewig, ``Improved online algorithms for knapsack
  and gap in the random order model,'' in \emph{Approximation, Randomization,
  and Combinatorial Optimization. Algorithms and Techniques (APPROX/RANDOM
  2019)}.\hskip 1em plus 0.5em minus 0.4em\relax Schloss
  Dagstuhl-Leibniz-Zentrum fuer Informatik, 2019.

\bibitem{mitrinovic1991inequalities}
D.~S. Mitrinovic, J.~Pecaric, and A.~M. Fink, \emph{Inequalities involving
  functions and their integrals and derivatives}.\hskip 1em plus 0.5em minus
  0.4em\relax Springer Science \& Business Media, 1991, vol.~53.

\end{thebibliography}

\end{document}